\documentclass[aps,pre,preprint,superscriptaddress]{revtex4-1}  
\usepackage{graphicx}
\usepackage{dcolumn}
\usepackage{amssymb}
\usepackage{amsmath}
\usepackage{newtxtext,newtxmath,lipsum}
\usepackage[shortlabels]{enumitem}

\newcommand{\sech}{\normalfont\mbox{sech}\,}
\newcommand{\csch}{\normalfont\mbox{csch}\,}

\begin{document}
	\title{Robustness of the solitons
against perturbations in certain nonlocal nonlinear Schr\"{o}dinger-type equations in Nonlinear Physics}
	\author{M.D. Sreelakshmi} 
	\affiliation{Department of Physics, Indian Institute of Science Education and Research, Tirupati 517 507, Andhra Pradesh, India}
	
	\author{N. Sinthuja}
	\affiliation{Department of Physics, Anna University, Chennai - 600 025, Tamilnadu, India}
	
	\author{N. Vishnu Priya}
	\affiliation{Department of Mathematics, Indian Institute of Science, Bangalore - 560012, Karnataka, India}
	
	\author{M. Senthilvelan}
	\affiliation{Department of Nonlinear Dynamics, Bharathidasan University, Tiruchirappalli - 620 024, Tamilnadu, India}
	\email{velan@cnld.bdu.ac.in}
	\begin{abstract}
		\par  The nonlocal nonlinear evolution equations describe phenomena in which wave evolution is influenced by local and nonlocal spatial and temporal variables. These equations have opened up a new wave of physically important nonlinear evolution equations.  Their solutions provide insights into the interplay between nonlinearity and nonlocality, making it a cornerstone in the study of nonlocal nonlinear systems.  However, the stability of such solutions has not been extensively explored in the literature. Stability analysis ensures that these solutions are robust and capable of persisting under real-world perturbations, making them physically meaningful.  In this work, we examine the stability of soliton solutions of four types of nonlocal nonlinear evolutionary equations: (i) the space-shifted nonlocal nonlinear Schr\"{o}dinger equation, (ii) the nonlocal complex time-reversed Hirota equation, (iii) the nonlocal real space-time-reversed modified Korteweg–de Vries equation, and (iv) a fourth-order nonlocal nonlinear Schr\"{o}dinger equation. These equations arise in various physical fields such as nonlinear optics, Bose-Einstein condensates, plasmas and so on where nonlocality and nonlinearity play significant roles. We introduce certain perturbations to the soliton solutions of these equations and analyze their stability.  Our findings indicate that the soliton solutions of the aforementioned equations are stable under such perturbations.  To the best of our knowledge, this approach to investigating the stability of these solutions is novel.
	\end{abstract}
	
	\maketitle
	\section{Introduction}
Nonlinear integrable systems have long been of great interest due to their exact solvability and the physically relevant solutions they provide \cite{Ablo}. These systems serve as idealized models for understanding complex nonlinear phenomena \cite{d1}.  Over the years, numerous discoveries have fueled rapid advancements in the study of integrable systems, leading to the development of other physically significant nonlinear wave equations \cite{clarkson}. Recent studies show that these equations can model real-world complex systems across various fields, including fluid dynamics,  plasma physics, nonlinear optics, and so on \cite{Jianke}. Among these the nonlinear Schr\"{o}dinger (NLS) equation stands as one of the well-studied and prototypical examples \cite{Fibich}. It  describes the slowly evolving wave packets across diverse nonlinear systems \cite{Balomed}.  The solutions of the NLS equation such as solitons, rogue waves and breathers constitute a captivating realm of nonlinear wave phenomena \cite{Kharif}.   
\\A decade ago, a nonlocal counterpart of NLS equation was introduced by Ablowitz and Musslimani of the form
\begin{equation}
	i\;q_t(x,t)+q_{xx}(x,t)+2\;q^2(x,t)\;q^\ast(-x,t)=0.
	\label{e1}
\end{equation}
which opened  up new avenues for the integrability community to study the nature and physical importance of various nonlocal nonlinear evolutionary equations \cite{Ablowitz}.  In Eq. (\ref{e1}), $q(x,t)$ represents the amplitude of the wave envelope which depends on the spatial variable $x$ and the temporal variable $t$. It is interesting to note that the complex conjugate of $q(x,t)$ depends on spatial varibale $-x$, which introduces a "nonlocal" feature. This means the equation involves terms that depend not only on the local values of the function but also on their values at other symmetrically related spatial points .  This nonlocal NLS equation (\ref{e1}) is completely integrable and invariant under the parity-time (PT) transformation, which means the equation remains unchanged under combined transformations of complex conjugation, $x\rightarrow-x$ and $t \rightarrow -t$. Its integrability has been established through the existence of a Lax pair, a mathematical structure that guarantees the system's solvability via the inverse scattering transform (IST). Additionally, the equation admits an infinite number of conservation laws, which reflect the preservation of specific quantities during the evolution  \cite{Ablowitz}.  Physical applications of Eq. (\ref{e1}) have been proposed in the fields of magnetics and optics \cite{{Agalarov},{Ablowitz2}}. {The seminal work by Ablowitz and Musslimani not only introduced Eq.(\ref{e1}) but also established a framework for the study of nonlocal integrable systems. Since then many  such equations including the nonlocal Hirota, nonlocal mKdV, reverse space-time nonlocal NLS, reverse space-time vector nonlocal NLS, complex reverse space-time nonlocal mKdV, real reverse space-time nonlocal mKdV, real reverse space-time nonlocal sine-Gordon, and reverse space-time nonlocal Davey-Stewartson, have been proposed and widely studied in the literature (for more such equations, refer to \cite{d1,Ablowitz2,{Fokas},{zhang},{AM2},{Yan},{Xin1},{Xin2}}.  The solutions of nonlocal equations, such as bright, dark, antidark solitons, breathers, rational solutions, and periodic solutions, have been well analyzed (see, for example, Refs. \cite{{Liming},{dad},{Ablowitz3},{chinese1},{chinese2},{Khare},{vishnu}}).  
\par However, the stability of these solutions has been very rarely investigated in the literature, with notable studies including the stability of soliton solutions against linear perturbations for the nonlocal Hirota equation and Alice and Bob versions of the Korteweg-de Vries and Boussinesq equations \cite{jcen}. They have shown that some of the solutions of these equations are unstable. Given the physical significance of nonlocal equations and the crucial role of stability in understanding the robustness of solutions, it is important that stability analysis should be extended to a wider range of systems. The key objective of this manuscript is to address this gap by investigating the stability properties of soliton solutions of certain integrable nonlocal equations.  For our investigation we consider the following four nonlinear evolution equations: (i) the space-shifted nonlocal NLS equation, (ii) the nonlocal complex time-reversed Hirota equation, (iii) the nonlocal real space-time reversed mKdV equation, and (iv) the fourth-order nonlocal NLS equation.  We show that the soliton solutions of the first three equations are stable.  The soliton solution of the fourth-order nonlocal NLS equation contains a singularity.  Hence, for our analysis, we consider the periodic solution of it and show that it is stable. To the best of our knowledge, the stability analysis of these equations has never been explored and we believe that our work will lay a strong foundation for future studies in this domain.  We also note here that, this work is primarily limited to the stability analysis of NLS-type and KdV-type nonlocal equations and focuses exclusively on small linear perturbations, which does not account for nonlinear interactions between perturbations and unstable solutions such as blow-up, collapse or transition to other states.  Furthermore, certain assumptions made during the derivation, such as specific symmetry properties and parameter choices, may restrict the applicability of the results to a broader class of systems.

Our work is structured as follows: In Section 2, we delve into the stability analysis of the space-shifted nonlocal NLS equation. In Section 3, we explore the stability properties of soliton solutions within the context of the complex time-reversed Hirota equation. Further, in Sections 4 and 5, we investigate the stability characteristics of the nonlocal real mKdV equation and the fourth-order nonlocal NLS equation, respectively. Finally, in Section 6, we provide a comprehensive summary of our findings and contributions.

\section{The space shifted nonlocal NLS equation}
A set of integrable reductions in the AKNS system led to a modified version of the nonlocal NLS equation that incorporates shifted terms in space, time, or both \cite{mjablowitz1}. In this section, we consider the following space-shifted PT-symmetric nonlocal reduction 
\begin{align}
    iq_t - q_{xx}-2\alpha q^2 q^\ast (k-x,t) = 0.
    \label{eq}
\end{align}
In Eq. (\ref{eq}) $q$ represents the slowly varying envelope field, the asterisk symbol denotes the complex conjugate and the subscripts stand for partial derivatives with respect to the respective variables. $\alpha= \pm1$, and $k$ is an arbitrary real parameter that represents the shift in the space. The interesting and remarkable aspect of such integrable systems is that the nonlocality occurs in a very different manner, because the field at a point $(x,t)$ is generally related to the field at the corresponding mirror-reflected and shifted point in the space domain, $(k-x,t)$. 
\par Here, we examine the linear stability of the soliton solution of Eq. (\ref{eq}) corresponding to $\alpha= 1$. Let $q_0(x,t)$ be the original solution.  We perturb the solution in the form
\begin{align}
    q(x,t) = q_0(x,t) + \epsilon   \sigma(x,t) ,
    \label{spaceshiftedeq2}
\end{align}
where $\sigma(x,t)$ is the perturbing function and $\epsilon <<1$. By substituting Eq. (\ref{spaceshiftedeq2}) into the original equation and neglecting the higher-order ($\ge 2$)  terms in $\epsilon$, we obtain
\begin{align}
 iq_{0t} -q_{0xx}-2q_0^2 q^\ast_0 (k-x,t)+  \epsilon[i\sigma_t-\sigma_{xx}-2q_0^2 \sigma^\ast (k-x,t)\nonumber\\-4\sigma q_0 q^\ast_0 (k-x,t)]=0.
 \label{subeq}
\end{align}
The above equation reduces to Eq. (\ref{eq}) in zeroth order of $\epsilon$ and vanishes.  Therefore, we need the first-order terms in $\epsilon$ to go to zero. 
\begin{align}
   i\sigma_t-\sigma_{xx}-2q_0^2 \sigma^\ast (k-x,t)-4\sigma q_0 q^\ast_0 (k-x,t)= 0.
   \label{perturbation eq}
\end{align}
This equation needs to be solved in order to find an appropriate perturbing function $\sigma(x,t)$ for the solution.
\subsection{Soliton solution}
The general form of the one soliton solution of Eq. (\ref{eq}), 
\begin{align}
    q(x,t) = \displaystyle\frac{-2(\eta+\eta_1) e^{-4i\eta^2 t} e^{-\eta k} e^{-i\theta} e^{2\eta x}}{1-e^{-(\eta+\eta_1)k}e^{-i(\theta+\theta_1)}e^{4i(\eta^2_1-\eta^2)t}e^{2(\eta_1+\eta)x}},
    \label{sol}
\end{align}
 has been derived in \cite{mjablowitz1} by Ablowitz and Musslimani using the inverse scattering transform.  This solution has four free parameters, of which $\eta$, $\eta_1$ are real and positive, and $\theta$ and $\theta_1$ are arbitrary real constants. We convert Eq. (\ref{sol}) into a simpler hyperbolic waveform in order to analyze the stability.  Upon setting $\eta_1= \eta$ and $\theta_1 = \theta$, we can rewrite the above solution in the form 
\begin{align}
    q(x,t) = \mu e^{-i\mu^2 t} \text{csch} (\mu x -\eta k -i\theta),
    \label{reducedsol}
\end{align}
with $\mu = 2\eta$. Compared to  the general solution, this simplified form reduces the number of free parameters to two, namely, $\eta$ and $\theta$  both of which are real. Additionally, it separates the spatial and temporal components in a more straightforward manner, making it easier to analyze its physical features.
\subsection{Stability of the solution}
\vspace{0.5cm}
\noindent \textbf{Theorem 1.} \textit{The one-soliton solution $q(x,t)$ of the space shifted nonlocal NLS equation is stable under both odd and even perturbations of the form $ ib \epsilon e^{-i\mu^2 t} \text{csch}(\mu x-\eta k -i\theta)$ and $ ib \epsilon e^{-i\mu^2 t} \text{csch}(\mu x-\eta k -i\theta) \text{coth}(\mu x-\eta k -i\theta)$ respectively, where b is a free parameter, $\eta$ and $\theta $ are arbitrary constants and $\epsilon \ll 1$, such that $b,\eta,\theta,\epsilon \in R$.} 
\vspace{0.5cm}
\\\\\textbf{Proof.} We allow the original solution and the perturbing function to take the following form:
\begin{align}
    q_0(x,t) = \mu e^{-i\mu^2 t} F[r(x)], \;
    \sigma(x,t) = i b e^{-i\mu^2 t} G[r(x)].
    \label{assumed pert1}
\end{align}
Similarly,
\begin{subequations}
\begin{eqnarray}
    q^\ast_0(k-x,t) = -\mu e^{i\mu^2 t} F[r(x)], \label{assumed sol2} \\ 
    \sigma^\ast(k-x,t) = -i b e^{i\mu^2 t} \omega G[r(x)].
    \label{assumed pert2}
\end{eqnarray}
\label{new}
\end{subequations}
We also assume $r(x)=\mu x -\eta k -i\theta $ and $\omega=\pm1$ for $G$ to be even or odd with respect to space-shifted parity conjugation. Now, by substituting Eqs. (\ref{assumed pert1}) and (\ref{new}) into (\ref{perturbation eq}), we obtain 
\begin{align}
    ibe^{-i\mu^2t} \mu^2 U_1=0,
    \label{pertub eq 2}
\end{align}
where
\begin{align}
 U_1 = G[r]+2(2+\omega) F[r]^2 G[r] - G^{(2)}[r].
 \label{aux1}
\end{align}
\par Here the superscript $(n)$ denotes the $n$th partial derivative with respect to $[r]$. Solving Eq. (\ref{aux1}) for $G[r]$ will give us the analytical form of the perturbing function. By comparing the solution (\ref{reducedsol}) with the assumed general form, we find that $F[r]$ can be identified as $\text{csch[r]}$. Now let us consider the two cases that arise here: $\omega= 1$ and $-1 $. For $\omega = 1$, Eq. (\ref{aux1}) becomes 
\begin{align}
    G[r]+ 6 \;\text{csch}^2[r] G[r] - G^{(2)}[r]= 0,
    \label{evencase}
\end{align}
On solving Eq. (\ref{evencase}), we get $G[r] = \text{csch}[r] \coth[r]$. This indicates that the solution to the first-order equation (\ref{perturbation eq}) is derived as follows: 
\begin{align}
    \sigma_n(x,t) = ib e^{-i\mu^2 t} \text{csch}(\mu x-\eta k -i\theta) \coth(\mu x-\eta k -i\theta),
    \label{evenpert1}
\end{align}
where the subscript $n$ denotes the $\omega = 1$ or even case. Similarly, for the next case,  Eq. (\ref{aux1}) becomes 
\begin{align}
    G[r]+ 2 \text{csch}^2[r] G[r] - G^{(2)}[r]= 0.
    \label{oddcase}
\end{align}
This equation is solved by $G[r] = \text{csch}[r]$ and therefore the perturbing function for this case takes the form,
\begin{align}
 \sigma_o(x,t) = ib e^{-i\mu^2 t} \text{csch}(\mu x-\eta k -i\theta),   
    \label{oddpert1}
\end{align}
where the subscript $o$ denotes the $\omega = -1$ or odd case.
The stability of the solution depends on how the perturbing function affects the original solution. If the perturbation solution $\sigma(x,t)$ introduces only deformations or oscillations to the original solution, then $q_0$ can be considered stable. When the perturbing function causes the solution to diverge when added to the original solution, it can be considered unstable. Both the perturbation functions found above, which is given in Eqs. (\ref{evenpert1}) and  (\ref{oddpert1}) vanishes as $x \rightarrow \pm \infty$. 

Figure \ref{fig1} shows the soliton solution for the nonlocal space-shifted NLS equation with perturbation for various values of $b$. We can see that odd perturbations only introduce minor variations to the solutions [Fig. \ref{fig1}(a)] and even perturbations introduce only small oscillations to them [Fig. \ref{fig1}(b)]. None of the perturbations leads to divergence of the solution. Hence, we conclude that the solution is stable .
\begin{figure}[!ht]
    \centering
    \includegraphics[width=\textwidth]{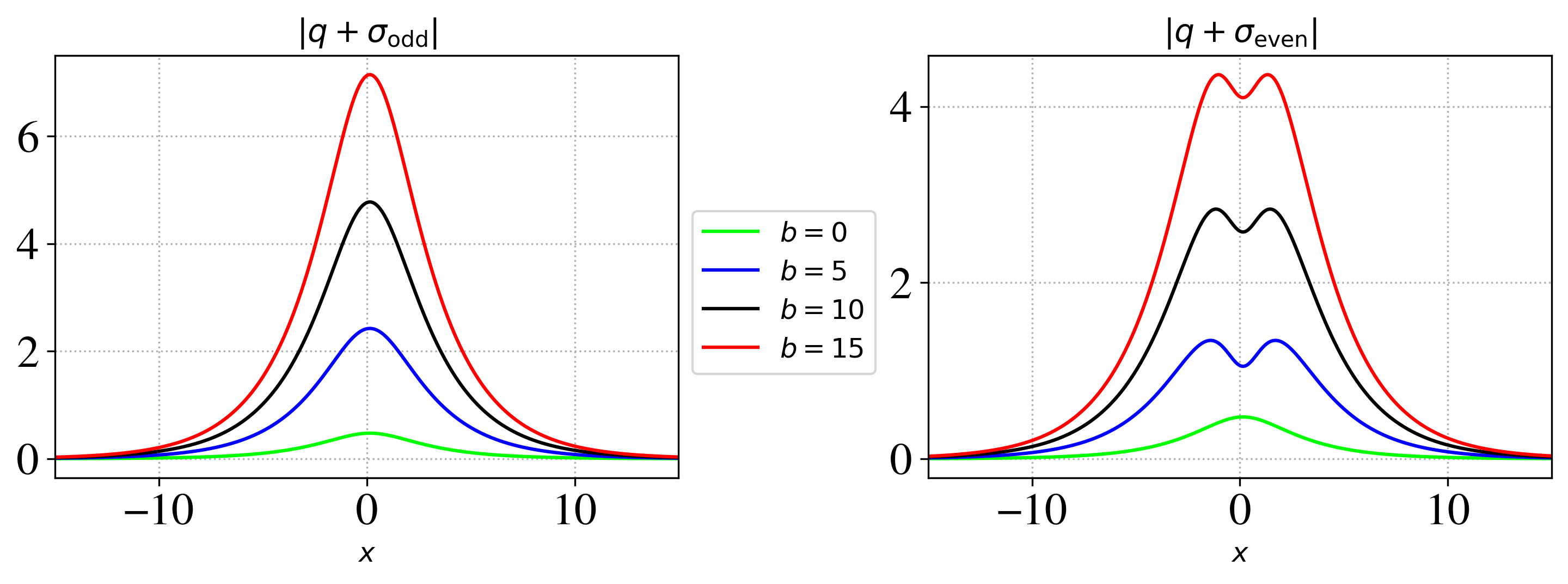}
    \caption{Soliton solutions for the space shifted nonlocal NLS equation (\ref{eq}) along with its (a) odd (Eq.\ref{oddpert1}) and (b) even (Eq.\ref{evenpert1}) perturbation for different values of $b$ and $t = 0$, $\epsilon= 0.4$, $\eta=0.2$, $\theta=1$, $k=0.3$. }
    \label{fig1}
\end{figure}
\section{The nonlocal complex time reversed Hirota equation}
In this section, we investigate the stability of the soliton solution of the nonlocal complex time reversed Hirota equation which is introduced in \cite{jcen2}.
\begin{align}
  iq_t + i\delta_1 (q_{xx} - 2kq^\ast(x,-t) q^2)-\delta(q_{xxx}-6kq q^\ast(x,-t) q_x) =0.
  \label{eq2}
\end{align}
where $q$ represents the wave profile at the point $(x,t)$,  $\delta_1, \delta \in \mathbb{R} $ and $k=\pm 1$. This is a third-order nonlocal NLS equation  with nonlocality in the time domain, that is, the function values at a point $(x,t)$ are related to the function values at $(x,-t)$.  $\delta$ is the parameter that decides the influence of higher-order effects. The stability analysis for the nonlocal Hirota equation in space was previously done in \cite{jcen}.  We present the analysis of the complex time-reversed Hirota equation here.

Again, replacing $q_0 \rightarrow q_0(x,t) + \epsilon\; \sigma(x,t)$ with $\epsilon << 1$, we obtain the first order equation in $\epsilon$ as 
\begin{align}
  i\sigma_t + i \delta_1 \sigma_{xx} -& i\delta_1 2k(2q_0 q^\ast_0(x,-t) \sigma + \sigma^\ast(x,-t) q_0^2)-\delta\sigma_{xxx}+ 6\delta k (q_0 q^\ast_0(x,-t) \sigma_x\nonumber\\&+ q_0 \sigma^\ast (x,-t) q_{0x}+\sigma q^\ast_0(x,-t) q_{0x}) = 0.
  \label{perturbing eq}
\end{align}
\subsection{Soliton Solution}
The solution for the nonlocal complex time-reversed equation has been derived by applying the Darboux-Crum transformations \cite{jcen2}. In this paper, we analyze the standard one soliton solution of this equation, which is given by 
\begin{align}
   q(x,t)= \displaystyle\frac{-2(\lambda+\lambda^\ast) e^{2\lambda x +\zeta_1 + \zeta^\ast_2}}{e^{2x(\lambda+\lambda^\ast)+4\lambda^{\ast 2} (\delta_1-2i\delta\lambda^\ast)t+\zeta_1+\zeta_1^\ast}+e^{4\lambda^2(\delta_1+2i\delta\lambda)t+\zeta_2+\zeta_2^\ast}} ,
    \label{sol2}
\end{align}
where $\lambda \in i\mathbb{R}$ and $\zeta_2=\zeta_1^\ast$. Using the transformation, $\lambda=\lambda^\ast$ and $\zeta_1=\zeta_2=0$, the above solution can be reduced to a simpler hyperbolic waveform 
\begin{align}
    q(x,t)=-\mu e^{-\mu^2\delta_1 t} \sech(\mu(x-i\delta \mu^2 t)).
    \label{reduced eq2}
\end{align}
where $\mu = 2\lambda$ and $k=-1$. Compared to the original solution, the simplified form eliminates the constants $\zeta_1$ and $\zeta_2$ while retaining $\delta$ and $\delta_1$ and uses a single real-valued free parameter $\mu$, further simplifying our analysis.
\subsection{Stability of the solution}
\vspace{0.5cm}
\noindent\textbf{Theorem 2.} \textit{The solution $q(x,t)$ of the reverse time Hirota equation remains stable under perturbations of the form $ i\nu \epsilon e^{-\mu^2 \delta_1 t} \sech[\mu(x-i\delta \mu^2 t)] $, where $\mu, \delta, \delta_1, \epsilon \in R$ and $\epsilon \ll 1$ with the perturbation analyzed as a function of time $t$.} 
\vspace{0.5cm}
\\\\\textbf{Proof.}
To solve Eq. (\ref{perturbing eq}) and to find the perturbing function, we define two new variables.
\begin{align}
    z_1 = x-i\delta \mu^2 t,\;  z_2 = x+ i\delta \mu^2 t.
\end{align}
\par We assume the solution and perturbing function to be of the form 
\begin{subequations}
 \begin{eqnarray}
     q_0(x,t)&=&\mu e^{-\mu^2\delta_1 t} F[\mu z_1(x,t)],\label{assumed 3a}\\
     \sigma(x,t)&= &i\nu e^{-\mu^2\delta_1 t} G[\mu z_1(x,t)], \label{assumed 3b}\\
     q^\ast_0(x,-t)&=&\mu e^{\mu^2\delta_1 t} F[\mu z_1(x,t)], \\
     \sigma^\ast(x,-t)&= &-i\nu e^{\mu^2\delta_1 t} G[\mu z_1(x,t)].\label{assumed 3d}
 \end{eqnarray}
 \end{subequations}
 Here, the perturbing function is assumed to be an even complex parity function. This assumption arises naturally due to the symmetry properties of the variable $z_1=x-i\delta\mu^2t$, which remains unchanged under the combined operations of time reversal $(t\rightarrow-t)$ and complex conjugation $(i\rightarrow -i)$ implying that $z_1$ is even under this transformation. Since the perturbing function $G$ is defined in terms of $z_1$, it is reasonable to assume that it also inherits this even symmetry. Substituting the assumed solutions and perturbations (\ref{assumed 3a})-(\ref{assumed 3d}) into Eq. (\ref{perturbing eq}) and neglecting the higher order terms in $\delta_1$ and $\delta$, we get 
 \begin{align}
     \nu e^{-i\frac{(z_1 - z_2)\delta_1}{2\delta}}\mu^2 [\delta_1 U_2 + i\delta U_3] =0,
     \end{align}
 where
 \begin{subequations}
 \begin{align}
     U_2 =&\; 2 k F[\mu z_1]^2 G[\mu z_1] + G[\mu z_1] - G^{(2)}[\mu z_1] ,\\
     U_3 = &\;\mu(F^{(1)}[\mu z_1]+6kF[\mu z_1]^2 G^{(1)}[\mu z_1]-G^{(3)}[\mu z_1]).
 \end{align}
 \end{subequations}
 \par For the above equation, we find that
 \begin{align}
\frac{\partial}{\partial z_1} (U_2) = U_3 + 4k \mu F[\mu z_1]\, W(G[\mu z_1],F[\mu z_1]),
 \label{aux eq}
 \end{align}
 where $W(G[\mu z_1],\;F[\mu z_1])= G[\mu z_1] F^{(1)}[\mu z_1]-F[\mu z_1] G^{(1)}[\mu z_1]$ is the Wronskian for the functions. From the relation between $U_2$ and $U_3$, we observe that if $G[\mu z_1]$ is set to be equal to   $F[\mu z_1] $, then the Wronskian becomes zero. Thus, if we can find a $F[\mu z_1]$ such that it satisfies 
 \begin{align}
 2kF[\mu z_1]^3 + F[\mu z_1]-F^{(2)}[\mu z_1] = 0,
 \label{aux2 eq }
 \end{align}
 then the first order equation in $\epsilon$ (Eq. (\ref{perturbing eq})) can be solved. By comparing the solution obtained from the equation and the assumed general form, we take $F[\mu z_1]=G[\mu z_1]=sech[\mu z_1]$. It can be verified that this assumption solves Eq. (\ref{aux2 eq }) which tells us that the perturbing function is given by
 \begin{align}
\sigma(x,t)= i\nu e^{-\mu^2 \delta_1 t} \sech[\mu(x-i\delta \mu^2 t)].
\label{reversetimepert}
 \end{align}
 \par Figure \ref{fig2} shows the solutions and the perturbed solutions for the reverse time Hirota equation at different times, it can be observed that the perturbations only cause very small variations in the solution and the amplitude of  the perturbed solution also starts decreasing after a time $t =5$. Therefore, we conclude that the solution is stable.
\begin{figure}[!ht]
    \centering
    \includegraphics[width=\textwidth]{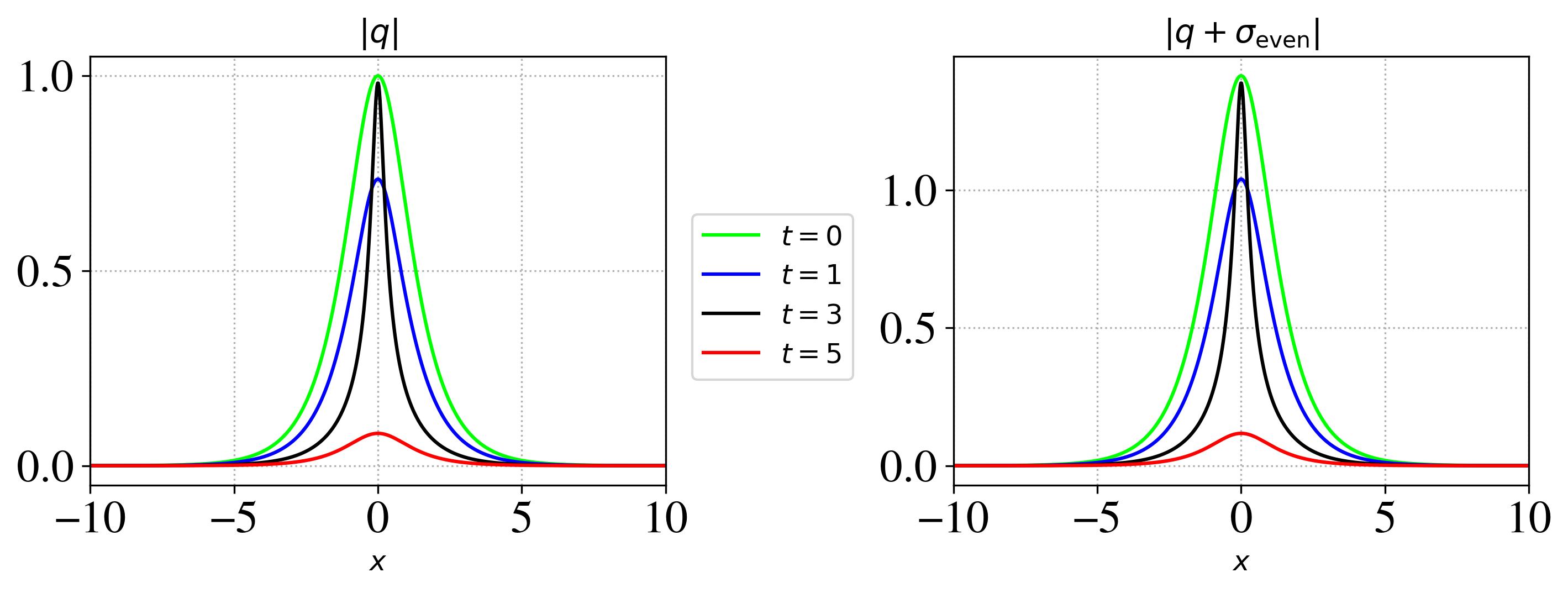}
    \caption{(a) Soliton solutions of the reverse time Hirota equation Eq.(\ref{reduced eq2})  and (b) perturbed solutions Eq.(\ref{reversetimepert}) for different times ($t$) with $\lambda=0.5$, $\delta_1=0.5$, $\delta=0.6$, $\nu = 2$ and $\epsilon=0.5$.}
    \label{fig2}
\end{figure}
 \section{The nonlocal real space-time reversed mKdV equation}
 Another important equation that belongs to the class of nonlocal equations is the nonlocal real space-time reversed mKdV equation \cite{Ablowitz2}
 \begin{align}
    q_t+ q_{xxx}-6\sigma qq(-x,-t)q_x = 0,
    \label{eq.3}
 \end{align}
where $q$ represents the wave profie and $\sigma = \pm 1$. This equation is nonlocal in both space and time, however, note that the field considered at $(-x,-t)$ is real, and not the complex conjugated. The stability of one soliton solution of the KdV equation and the Alice and Bob KdV system has been studied previously \cite{jcen}. Here, we investigate the stability of the soliton solution of Eq. (\ref{eq.3}) for the $\sigma = -1$ case. Replacing the original solution $q_0(x,t)$ with $q_0(x,t)+ \epsilon \sigma(x,t)$ ($\epsilon <<1$) and equating the terms of the first order in $\epsilon$ to zero, we arrive at the equation for the perturbing function in the form
\begin{align}
   \sigma_t + \sigma_{xxx}+6[q q(-x,-t) \sigma_x + q \sigma(-x,-t) q_x + \sigma q(-x,-t) q_x ]=0.
   \label{pert eq3}
\end{align}
\subsection{One soliton solution}
The derivation of the one-soliton solution for nonlocal mKdV equation can be found in \cite{Ablowitz2} The general expression delineating the one-soliton solution within the framework of real space-time nonlocal equation (\ref{eq.3}) is given by
\begin{align}
    q(x,t)=\frac{2\gamma_1 (\eta _1 + \eta _2) e^{-2\eta_2 x+8 \eta^3_2 t}}{1+\gamma_2 e^{-2 \eta_1 x+8\eta^3_1 t-2\eta_2 x+8 \eta^3_2 t}},
\end{align}
where $\gamma_1,\gamma_2 = \pm 1$ , $\eta_1$ and $\eta_2 $ are  free real parameters. In the context of present discussion, when $\gamma_1 = \gamma_2 = 1$ is taken into account, the resultant solution can be aptly represented in the form:
\begin{align}
    q(x,t) =\frac{ (\eta _1 + \eta _2) e^{\eta_1(x-4\eta^2_1 t)} e^{-\eta_2(x-4 \eta^2_2 t)}}{\cosh[\eta_1(x-4\eta^2_1 t)+\eta_2(x-4 \eta^2_2 t)]}.
\end{align}
This solution is non-singular. Again by setting $\eta_1 = \eta_2 =\eta$, the solution can be further simplified to the form
\begin{align}
    q(x,t) =\mu \sech[\mu(x-\mu^2 t)], \label{eqnew1}
\end{align}
where $\mu = 2\eta$. The simplified solution, reduces to a hyperbolic-type solution that depends on a single real parameter $\mu$ and we get a compact form compared to the general solution.
\subsection{Stability analysis}
\vspace{0.5cm}
\noindent\textbf{Theorem 3.} \textit{The one-soliton solution $q(x,t)$ of the nonlocal real space-time reversed mKdV equation is stable under both odd and even perturbations of the form $ ib \epsilon\sech[\mu(x-\mu^2 t)] \tanh[\mu(x-\mu^2 t)]$ and $ ib\epsilon \sech[\mu(x-\mu^2 t)]$ respectively, where b is a free parameter, $\mu$ is an arbitrary constant and $\epsilon \ll 1$, such that $b,\mu,\epsilon \in R$.} 
\vspace{0.5cm}
\\\\\textbf{Proof.}
To prove the stability of the one soliton solution (\ref{eqnew1}), we define two variables in order to replace $x$ and $t$, 
\begin{align}
    z_1 = & x-\mu^2 t, \;    z_2 = x+\mu^2 t.
\end{align}
Moving forward, we consider the soliton solution and perturbing function to exhibit the subsequent form:
\begin{subequations}
	\begin{align}
    q(x,t) =& \;\mu F[\mu z_1] \label{assume sol 4a},\; q(-x,-t)= \;\mu F[\mu z_1],\\
    \sigma(x,t)=& \;ib G[\mu z_1],\; \sigma(-x,-t)= \;-ib \omega G[\mu z_1].
    \end{align}
\label{assume pert 4d}
\end{subequations}
 Here also $\omega = \pm 1$ represents the even and odd parity-time symmetries respectively. By substituting all the above assumptions in (\ref{pert eq3}), we arrive at 
 \begin{align}
    ib \mu^3(G^{(3)}[\mu(z_1)]-(1-6 F[\mu z_1]^2)G^{(1)}[\mu z_1)]\nonumber\\+6(1-\omega) F[\mu z_1] F^{(1)}[\mu z_1]G[\mu z_1]) =0.
     \label{auxeq3}
 \end{align}
 \par Again, by comparing our solution with the assumed form, we observe that $F[\mu z_1]= \sech[\mu z_1]$. Now, we consider two separate cases in our discussion. For $\omega=1$, Eq. (\ref{auxeq3}) becomes
 \begin{align}
      ib\mu^3(G^{(3)}[\mu(z_1)]-(1-6\sech[\mu z_1]^2)G^{(1)}[\mu z_1)])=0.
      \label{even case2}
 \end{align}
 We find that Eq. (\ref{even case2}) is solved by $G[\mu z_1]= \sech[\mu z_1]$. Therefore the perturbing function in this case is 
 \begin{align}
     \sigma_e (x,t) = ib \sech[\mu(x-\mu^2 t)],
     \label{pert func mkdv}
 \end{align}
 Similarly for $\omega=-1$, the equation becomes
 \begin{align}
      ib\mu^3(G^{(3)}[\mu(z_1)]-(1-6 \sech[\mu z_1]^2)G^{(1)}[\mu z_1]\nonumber\\+12 \sech[\mu z_1] \sech^{(1)}[\mu z_1]G[\mu z_1]) = 0.
      \label{odd case3}
 \end{align}
 Equation (\ref{odd case3}) is solved by $G[\mu z_1]= \sech[\mu z_1] \tanh[\mu z_1]$. So the perturbing function is 
 \begin{align}
     \sigma_o(x,t) =ib \sech[\mu(x-\mu^2 t)] \tanh[\mu(x-\mu^2 t)].
     \label{odd pert func mkdv}
 \end{align}
 \par Figure \ref{fig3} shows the solution of the nonlocal real space-time reversed mKdV equation along with its odd and even perturbations for various values of $b$. Here, we observe that the odd perturbations only introduce oscillations to the solution (Fig. \ref{fig3}(a)) while the even perturbations result in minor deformations (Fig. \ref{fig3}(b)). Therefore, the solution is stable.
\begin{figure}[!ht]
    \centering
    \includegraphics[width=\textwidth]{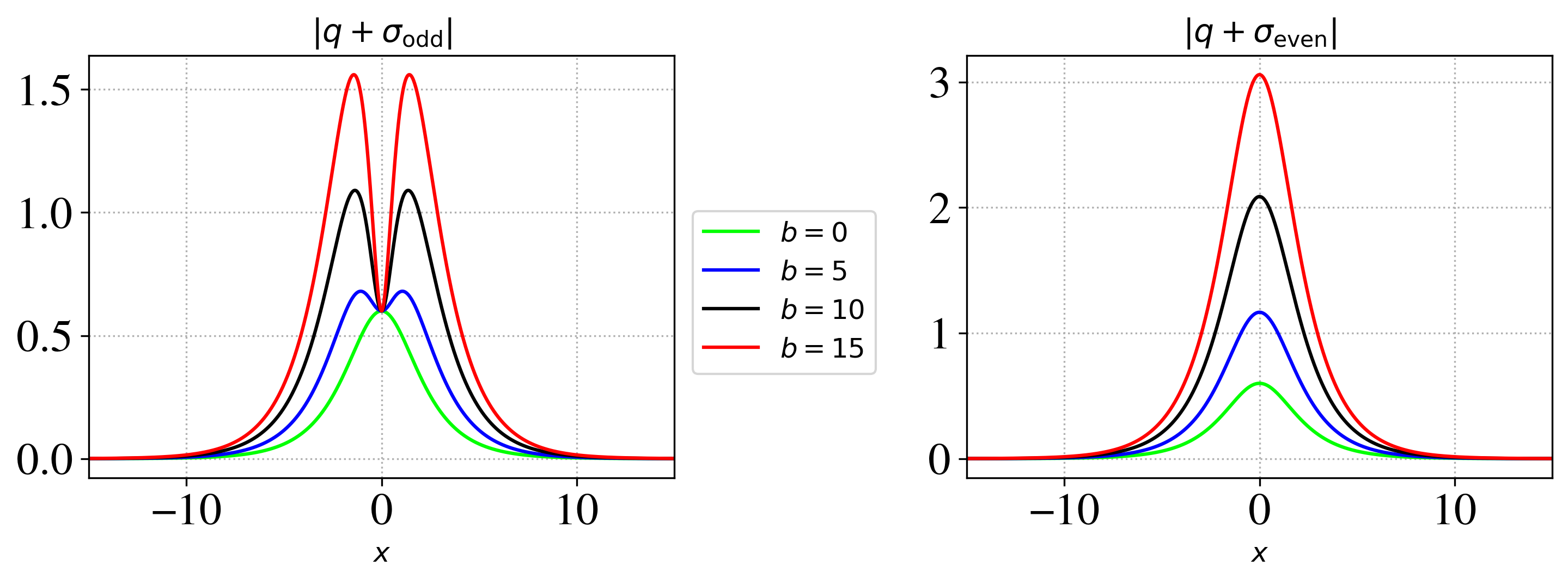}
    \caption{Soliton solutions for the nonlocal real space-time reversed mKdV equation with its (a) odd (Eq. \ref{odd pert func mkdv}) and (b) even (Eq. \ref{pert func mkdv}) for various values of $b$ and  $t= 0$, $\epsilon=0.2$, $\eta = 0.3$. }
    \label{fig3}
\end{figure}
\section{The fourth order nonlocal NLS equation}
Finally, we consider a fourth order nonlocal NLS equation \cite{radha}
\begin{align}
    iq_t+q_{xx}+2\mu q^2 \hat{q}+\delta[\mu(q_{xxxx}+6q^3 \hat{q}^2)+2q^2 \hat{q}_{xx}+6 q_x ^2 \hat{q}+4q q_x \hat{q}_x+8q \hat{q} q_{xx}]=0,
    \label{lpd}
\end{align}
where $q$ represents the wave envelope, $\hat{q}$ represents $q^\ast(-x,t)$, and $\mu$ and $\delta$ are real parameters. This equation is a higher-order NLS type equation with fourth-order dispersion, second-order dispersion, and cubic and quintic nonlinearities. $\delta$ is the parameter that decides the influence of higher-order effects. As $\delta$  tends to zero, equation (\ref{lpd}) reduces to the nonlocal NLS equation.  Here we focus on studying the stability of solutions of this fourth-order nonlocal NLS equation (\ref{lpd}) against linear perturbations. Let $q_0(x,t)$ be the original solution which is replaced by $q(x,t)=q_0(x,t)+\epsilon \sigma (x,t)$ where $\sigma (x,t)$ is the perturbing function and $\epsilon << 1$. By substituting this expression in the original equation, and neglecting higher-order ($\geq$2) terms in $\epsilon$, we obtain
\begin{align}
    iq_{0t}&+q_{0xx}+2\mu q_0^2 \hat{q}_0+\delta[\mu(q_{0xxxx}+6q_0^3 \hat{q_0}^2)+2q_0^2 \hat{q}_{0xx}+6 q_{0x}^2 \hat{q}_0+4q_0 q_{0x} \hat{q}_{0x}\nonumber\\&+8q_0 \hat{q}_0q_{0xx}]+\epsilon[i \sigma_t +\sigma_{xx}+2\mu q_0^2 \hat{\sigma}+4\mu q_0 \hat{q}_0 \sigma+\delta[\mu \sigma_{xxxx}+ 12\mu q_0^3 \hat{q}_0 \hat{\sigma}\nonumber\\&+ 18\mu q_0^2 \hat{q}_0^2 \sigma+2q_0^2 \hat{\sigma}_{xx}+4q_0  \hat{q}_{0xx} \sigma+6q_{0x}^2 \hat{\sigma}+12q_{0x} \hat{q}_{0} \sigma_x+4q_0 q_{0x} \hat{\sigma}_{x}+ 4q_0 \hat{q}_0 \sigma_x\nonumber\\&+4\sigma q_{0x} \hat{q}_{0x}+8q_0 \hat{q}_0 \sigma_{xx}+8q_0  \hat{\sigma} q_{0xx}+8\sigma \hat{q}_0 q_{0xx}]]=0.
\end{align} \label{eq1.2}

The above equation reduces to (\ref{lpd}) for $q_0$ in zeroth order and vanishes. Therefore, by equating the first-order terms in  $\epsilon$ to go to zero, we get
\begin{align}
    i \sigma_t& +\sigma_{xx}+2\mu q_0^2 \hat{\sigma}+4\mu q_0 \hat{q}_0 \sigma+\delta[\mu \sigma_{xxxx}+ 12\mu q_0^3 \hat{q}_0 \hat{\sigma}+ 18\mu q_0^2 \hat{q}_0^2 \sigma+2q_0^2 \hat{\sigma}_{xx}\nonumber\\&+4q_0  \hat{q}_{0xx} \sigma+6q_{0x}^2 \hat{\sigma}+12q_{0x} \hat{q}_{0} \sigma_x+4q_0 q_{0x} \hat{\sigma}_{x}+ 4q_0 \hat{q}_{0x}\sigma_x+4\sigma q_{0x} \hat{q}_{0x}\nonumber\\&+8q_0 \hat{q}_0 \sigma_{xx}+8q_0  \hat{\sigma} q_{0xx}+8\sigma \hat{q}_0 q_{0xx}]=0.
    \label{lpdperteq}
\end{align}
\subsection{One soliton solution}
The one soliton solution for Eq. (\ref{lpd}) has been derived using the Darboux transformation \cite{radha}. The solution is given by
\begin{align}
    q(x,t) = \frac{4i\mu a b^\ast \lambda_R}{|a|^2 e^{-2i\lambda^\ast x-4it\beta^\ast}-\mu |b|^2 e^{2i\lambda x-4it\beta}},
    \label{lpdsol}
\end{align}
here, $a, \;b,\; \lambda\in \mathbb{C},\;\mu \in \mathbb{R}, \;\beta= -\lambda^2 + 4\eta\lambda^4$ and $\eta = \mu \delta$ with the subscripts $R$ and $I$ denoting the real and imaginary parts of the respective variables. This solution exhibits different qualitative behaviors for different values of $\mu,\; \delta,\; \lambda$. In our analysis, we consider the non-singular behaviour which is possible if we set $\lambda_I = 0$ and $|a/b| \neq 1$. Our aim is to convert Eq. (\ref{lpdsol}) into a hyperbolic waveform. The equation can be reduced to the form of a stationary periodic solution, which is given by  
\begin{align}
    q(x,t) = 2i \frac{a}{b} \lambda_R e^{-\nu/2+4it\beta}\text{csch}(\nu/2-2i\lambda_R x),
    \label{lpdredsol1}
\end{align}
where $\nu=\ln\frac{a^2}{b^2}$. Again, by setting $2 \lambda_R =1$, and $a_I = b_I = 0$, we simplify the solution further to obtain 
\begin{align}
    q(x) = i\frac{a}{b}  e^{-\nu/2} \text{csch}(\nu/2-i x).
\end{align}
 Compared to the original solution, the simplified form eliminates the time-dependence, leaving the solution as a hyperbolic waveform in space. The solution now depends on the real parameter $\nu$, which is derived from the ratio of the parameters $a$ and $b$, and the spatial coordinate $x$. Moreover, the non-singularity condition is ensured in the simplified form, making it easier to analyze.
\subsection{Stability analysis}
\vspace{0.5cm}
\noindent\textbf{Theorem 4.} \textit{The one-soliton solution $q(x,t)$ of the fourth-order nonlocal NLS equation is stable under  perturbations of the form $ ik\epsilon e^{-\nu/2}\csch(\nu/2-ix)$ , where k is a free parameter, $\nu $ is an arbitrary constant and $\epsilon \ll 1$, such that $k, \nu,\epsilon \in R$.} 
\vspace{0.5cm}
\\\\\textbf{Proof.}
In order to find out the perturbing function which can tell us if the solution is stable or not, we first assume a form for the original solution given by 
\begin{align}
    q_0(x) = i\frac{a}{b}  e^{-\nu/2} F[r(x)].
\end{align}
Similarly, the perturbing function is assumed to be of the form
\begin{align}
    \sigma(x)= k  e^{-\nu/2} G[r(x)].
\end{align}
We also assume $r(x) = (\nu/2-i x) $. The parity conjugates of the above functions are assumed to be 
\begin{align}
    q^\ast_o (-x) = &-i\frac{a}{b} e^{-\nu/2} F[r(x)],\nonumber\\
    \sigma^\ast(-x) = & k e^{-\nu/2} G[r(x)].
\end{align}
 Here, we assume the perturbing function to be an even complex parity function since the solution has a form $r(x)$ that is invariant under the operations $x\rightarrow -x$ and $i\rightarrow -i$. This symmetry is inherent in the form of the solution. The perturbation equation is then expressed in terms of this function, ensuring that the perturbing term maintains this symmetry. By substituting the above assumed forms in Eq. (\ref{lpdperteq}), we obtain
\begin{align}
\frac{ke^{-5\nu/2}}{b^4}&(6a^4 F[r]^4 G[r]+2a^2b^2 e^\nu(G[r]F^{(1)}[r]^2-2F[r](3F^{(1)}[r]G^{(1)}[r]\nonumber\\&+G[r]F^{(2)}[r])F[r]^2(G[r]-3G^{(2)}[r]))-b^4e^{2\nu}(G^{(2)}[r]-G^{(4)}[r])=0.  \label{lpdsubstituted pert eq}
    \end{align}
Since $\nu = ln\frac{a^2}{b^2}$, we can find  
  \begin{align}  
 e^\nu =  \frac{a^2}{b^2}, \quad
 e^{2\nu} =  \frac{a^4}{b^4}. 
 \label{13}
\end{align}
\par Substituting the expressions (\ref{13}) in Eq. (\ref{lpdsubstituted pert eq}), we find 
\begin{align}
 \frac{ke^{-5\nu/2} a^4}{b^4}& (6F[r]^4 G[r]+2G[r]F^{(1)}[r]^2-12F[r]3F^{(1)}[r]G^{(1)}[r]-4F[r]G[r]F^{(2)}[r]\nonumber\\&+2F[r]^2G[r]-6F[r]^2G^{(2)}[r]-G^{(2)}[r]+G^{(4)}[r]) = 0.
 \label{final pert eq}
\end{align}
Solving the above equation for $G[r]$ will give us the analytical form of the equation. By comparing the assumed form of the solution with the solution derived in the previous section, we find that $F[r] = \text{csch}(r)$. Substituting for $F[r]$ we find that the solution to the above equation is $G[r]=\text{csch}[r]$. The perturbation function turns out to be
\begin{align}
    \sigma(x) = ke^{-\nu/2}\text{csch}(\nu/2-ix).
\end{align}
\par Figure \ref{lpd fig.} shows the solutions of the fourth order nonlocal NLS equation with its perturbations for values of $k$ in the perturbation. We observe that the perturbations only change the amplitude of oscillations of the solutions.  Therefore, we conclude that the solution is stable. 
\begin{figure}[!ht]
\begin{center}
			\includegraphics[width=8cm, height=6cm]{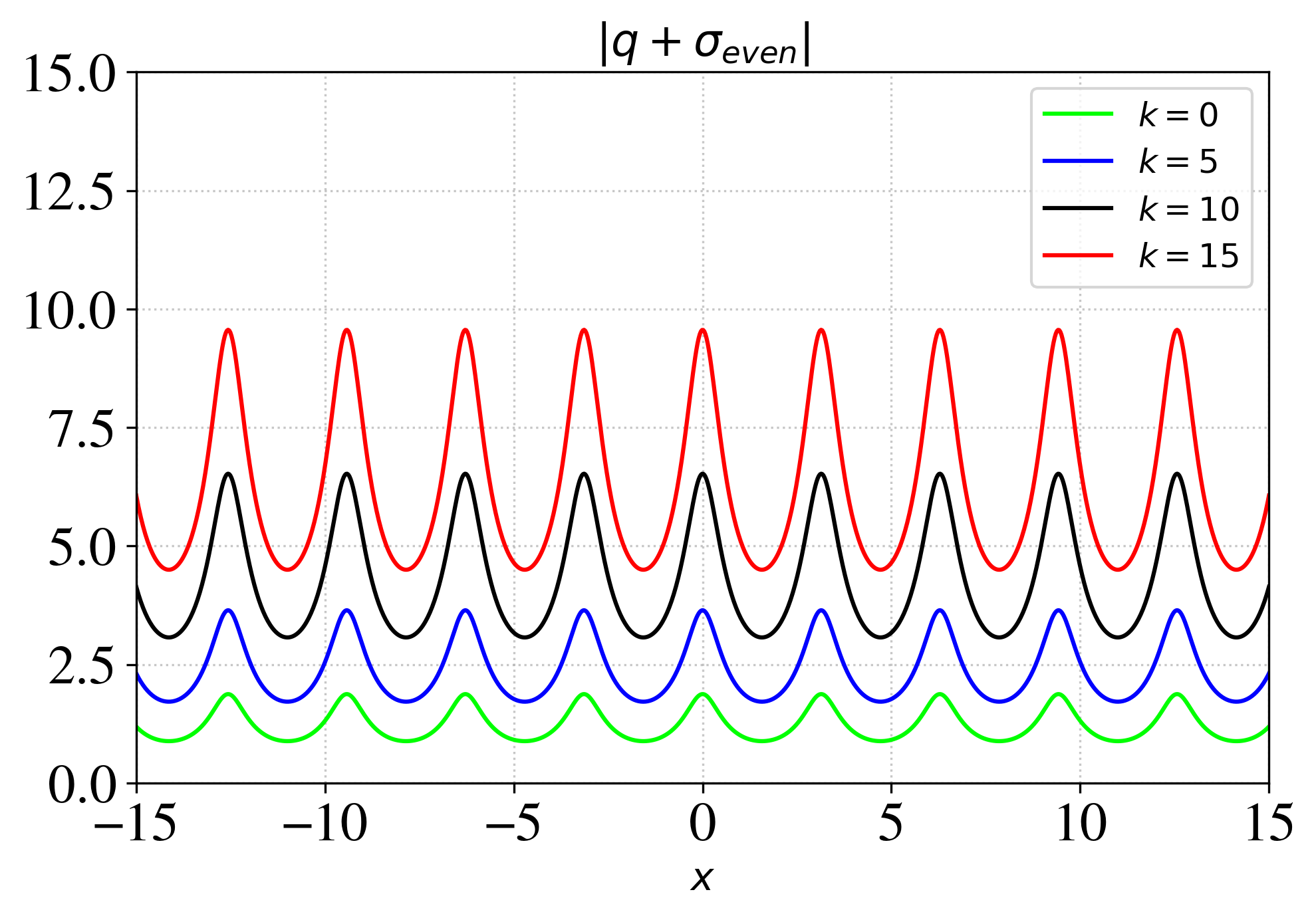}
   \end{center}
			\caption{Perturbed soliton solution for the fourth order nonlocal NLS equation for values of $k$ and $t=0$, $\epsilon=0.2,a=0.3,b=0.5$.}
              \label{lpd fig.}
\end{figure}
\section{Conclusion}
We investigated the stability of one soliton solutions for four nonlinear nonlocal evolutionary equations, namely (i) space shifted nonlocal NLS equation, (ii) complex time reversed Hirota equation, (iii) real modified KdV equation and (iv) fourth-order nonlocal NLS equation. For the nonlocal version of the space-shifted NLS equation, we discovered the small oscillation form of the solution and the corresponding perturbation terms, $\sigma$ (both odd and even perturbations), derived from the known one-soliton solution. Using this solution along with the respective perturbations, we found that the solutions are stable. Here, the space-shifted term $k$ and the perturbing function $\sigma$ play an important role in the dynamical changes. The same approach has been extended to complex reverse time Hirota equation. In that case as well, our investigations confirm that the solutions are stable. We also extended our study to the real reverse space-time mKdV and fourth-order nonlocal NLS equations and found that the solutions are stable. Our results will be valuable for understanding how the nonlocal systems evolve when subjected to linear perturbations. Nonlocal equations provide a means to describe how waves move through materials with varying properties, considering interactions between different points. The solutions of nonlocal equations may suggest potential applications in systems with nonlocal interactions or systems with PT-symmetry such as liquid crystal fibres, dipole-dipole interactions in BECs, long-range interactions in plasma waves such as Langmuir turbulence. In this context, stability analysis serves as a tool to assess whether the shapes of these waves, known as wave profiles, will remain bounded and controlled as they evolve over time. This analysis is crucial for predicting the behavior of signals and waves in practical scenarios, such as engineering designs and environmental predictions, ensuring that small disturbances do not lead to unpredictable or detrimental outcomes. All the models studied here exhibit different manifestations of nonlocality, each showcasing distinct behaviors. For example, \cite{jcen2} introduces several new integrable systems corresponding to nonlocal versions of the Hirota equation. The stability of these systems can be analyzed using a similar approach as presented in this work. Additionally, this methodology can be extended to higher-order versions of these equations. There are interesting open questions for further research. These potential directions offer the opportunity to deepen our understanding of nonlocal equations.


\begin{thebibliography}{90}
\bibitem{Ablo} Ablowitz M 2011 Nonlinear Dispersive Waves: Asymptotic Analysis and Solitons (New York:Cambridge University Press)

\bibitem{d1} Christodoulides D and Yang J 2018 Parity-time Symmetry and Its Applications (Singapore:Springer)

\bibitem{clarkson} Ablowitz M J and Clarkson P A 1991 Solitons, Nonlinear Evolution Equations and Inverse Scattering (Cambridge:Cambridge University Press)

\bibitem{Jianke} Yang J 2010 Nonlinear Waves in Integrable and Nonintegrable Systems (Philadelphia:SIAM)

\bibitem{Fibich} Fibich G 2015 The Nonlinear Schr\"{o}dinger Equation: Singular Solutions and Optical Collapse (Switzerland:Springer)

\bibitem{Balomed} Malomed B 2005 Nonlinear Schr\"{o}dinger Equation in: Scott A (ed.) Encyclopedia of Nonlinear Science (London:Routledge)

\bibitem{Kharif} Kharif C, Pelinovsky E, and Slunyaev A 2009 Rogue Waves in the Ocean: Observation, Theories and Modeling (New York:Springer)

\bibitem{Ablowitz} Ablowitz M J and Musslimani Z H 2013 Phys. Rev. Lett. \textbf{110} 064105

\bibitem{Agalarov} Gadzhimuradov T A and Agalarov A M 2016 Phys. Rev. A \textbf{93} 062124

\bibitem{Ablowitz2} Ablowitz M J and Musslimani Z H 2016 Stud. Appl. Math. \textbf{139} 7

\bibitem{Fokas} Fokas A S 2016 Nonlinearity \textbf{29} 319

\bibitem{zhang} Zhang Y S, Qiu D Q, Cheng Y and He J S 2017 Rom. J. Phys. \textbf{62} 108

\bibitem{AM2} Ablowitz M J, Luo X D and Musslimani Z H 2018 J. Math. Phys., \textbf{59} 011501

\bibitem{Yan} Yan Z 2015 Appl. Math. Lett. \textbf{47} 61

\bibitem{Xin1} Gao X Y 2024 Chin. J. Phys. \textbf{92} 1233-1239

\bibitem{Xin2} Gao X Y 2025 Appl. Math. Lett. \textbf{159} 109262

\bibitem{Liming} Huang X and Ling L 2016 Eur. Phys. J. Plus \textbf{131} 148

\bibitem{dad} Li M and Xu T 2015 Phys. Rev. E \textbf{91} 033202

\bibitem{Ablowitz3} Ablowitz M J and Musslimani Z H 2016 Nonlinearity \textbf{29} 915

\bibitem{chinese1} Li M, Xu T and Meng D 2016 J. Phys. Soc. Jpn. \textbf{85} 124001

\bibitem{chinese2} Wen X Y, Yan Z and Yang Y 2016 Chaos \textbf{26} 063123

\bibitem{Khare} Khare A and Saxena A 2015 J. Math. Phys. \textbf{56} 032104

\bibitem{vishnu} Vishnu Priya N, Senthilvelan M and Rangarajan G 2019 Phys. Lett. A \textbf{383} 15-26

\bibitem{jcen} Cen J, Correa F, Fring A and Taira T 2022 Phys. Lett. A \textbf{435} 128060

\bibitem{mjablowitz1} Ablowitz M J and Musslimani Z H 2021 Phys. Lett. A \textbf{409} 127516

\bibitem{jcen2} Cen J, Correa F and Fring A 2019 J. Math. Phys. \textbf{60} 081508

\bibitem{radha} Gadzhimuradov T A, Alagorov A M, Radha R and Tamil Arasan B 2020 Nonlinear Dyn. \textbf{99} 1295-1300
	\end{thebibliography}
\end{document}